\documentclass[conference]{IEEEtran}
\ifCLASSINFOpdf
\else
\fi
\usepackage{url}
\usepackage{cite}
\usepackage{amsmath}
\usepackage{amssymb}
\usepackage{breqn}
\usepackage[utf8]{inputenc}
\usepackage[T1]{fontenc}
\usepackage{textcomp}
\usepackage{gensymb}
\usepackage{upquote}
\usepackage{tabulary}
\usepackage{booktabs}
\usepackage{color,soul}
\usepackage{xcolor}
\usepackage{amsthm}        
\usepackage{graphics}
\usepackage{mathtools}
\usepackage{algorithm}
\usepackage{algpseudocode}
\usepackage{amsmath}

\usepackage[top=1in,bottom=1in,left=0.7in,right=0.7in,includefoot]{geometry}

\usepackage{pifont} 
\newcommand{\cmark}{\ding{51}}%
\newcommand{\xmark}{\ding{55}}%

\usepackage{algorithm}
\usepackage{algpseudocode}

\begin{document}
\title{Network Slice Embedding over Space Division Multiplexed Elastic Optical Networks}

\author{\IEEEauthorblockN{Divya Khanure \textsuperscript{$*$}, Riti Gour\textsuperscript{$\dagger$}, Congzhou Li\textsuperscript{$*$}, and Jason P. Jue\textsuperscript{$*$} }

\IEEEauthorblockA{\textsuperscript{$*$}Department of Computer Science\\
The University of Texas at Dallas, Richardson, Texas 75080, USA\\
\textsuperscript{$\dagger$} Department of Aviation and Technology,\\ San Jose State University, San Jose, California 95192, USA
}\\
\vspace{-0.5 cm}
Email:\{divya.khanure, jjue, congzhou.li\}@utdallas.edu, riti.gour@sjsu.edu}

\maketitle

\begin{abstract}
Network slicing over space-division-multiplexed elastic optical networks (SDM-EONs) enables efficient multiservice provisioning on a shared optical substrate. However, embedding such slices requires coordinated spectrum and compute resource management under dynamic traffic, which most existing RMCSA studies treat independently. This paper focuses on the network slice embedding problem over space-division-multiplexed elastic optical networks (SDM-EONs), aiming to develop efficient resource allocation strategies that ensure both high utilization and reliable service performance. While prior studies have investigated routing, modulation format, core, and spectrum allocation (RMCSA), they typically consider these dimensions separately from compute placement. To address this gap, this paper proposes a Waypoint-Assisted Multi-Segment Slice Mapping (WMSM) scheme, which integrates compute placement with spectrum allocation in a sequential but coupled manner to enable more flexible resource placement and improved spectrum efficiency. Numerical results show that WMSM improves acceptance ratios by up to 27\% under high-load conditions, while achieving up to 47\% lower total provisioning cost relative to the baseline strategy. These results highlight the benefits of integrated compute–spectrum provisioning and provide design insights for scalable, compute-aware optical slice mapping.
\end{abstract}


\IEEEpeerreviewmaketitle

\section{Introduction}

The exponential growth of bandwidth-intensive and cloud-driven applications continues to push the limits of optical transport infrastructures. Space-division multiplexing (SDM) combined with elastic optical networks (EONs) has emerged as a promising solution, enabling multiple spatial cores per fiber and dynamic spectrum allocation to deliver substantial improvements in capacity, flexibility, and spectral efficiency~\cite{ChatterjeeTutorial}.

Network slicing further enhances this flexibility by allowing multiple virtual networks to coexist on shared physical resources. In SDM-EONs, slicing spans three dimensions: spectrum, spatial cores, and computational capacity. This introduces coupled constraints — spectrum assignment must satisfy continuity and contiguity, and cores must be selected to mitigate inter-core crosstalk, and compute placement must respect node capacity limits — making joint optimization computationally intractable, while independent treatment leads to poor utilization and increased blocking.

Prior studies on network slicing over EONs typically treat compute placement separately from routing, modulation, core, and spectrum assignment (RMCSA), often leading to suboptimal or infeasible slice mappings: spectrum may be reserved along a path where compute capacity is unavailable, or vice versa. Most optical slice-embedding works also assume a strict one-to-one mapping between virtual and substrate nodes, which limits utilization. In the broader VNE literature, Wang~\textit{et al.}~\cite{NodeFusion} and Chowdhury~\textit{et al.}~\cite{ChowdhuryCoordinatedVNE} showed that node co-location and coordinated node-link mapping improve embedding efficiency and acceptance ratio, yet such strategies have not been extended to optical or SDM-EON slicing frameworks. Our work introduces controlled co-location of slice nodes on shared substrate nodes, improving compute utilization and reducing provisioning cost.

To address these gaps, this paper proposes two integrated heuristics for slice mapping in SDM-EONs. \textit{Direct Path-Driven Slice Mapping} (DPSM) serves as a baseline, prioritizing direct end-to-end routing with sequential compute placement. 
\textit{Waypoint-Assisted Multi-Segment Slice Mapping} (WMSM) introduces intermediate compute-capable waypoints and performs segment-wise spectrum allocation, enabling flexible, resource-efficient multi-segment provisioning. Both algorithms couple compute placement and spectrum allocation in a sequential but interdependent manner.

The main contributions are:
\begin{itemize}
    \item A coordinated system model for SDM-EON slice mapping that integrates spectrum, compute, and inter-core crosstalk constraints in a sequential but interdependent manner.
    \item A WMSM algorithm evaluated against a DPSM baseline, both with and without balanced placement variants.
    \item A performance evaluation demonstrating significant reductions in blocking probability, spectrum usage, and provisioning cost over baseline prioritization schemes.
\end{itemize}

The rest of the paper is organized as follows: Section~II reviews related work. Section~III presents the system model and problem formulation. Section~IV describes the proposed algorithms. Section~V evaluates performance, and Section~VI concludes.

\begin{figure*}
\centering
\includegraphics[width=\linewidth]{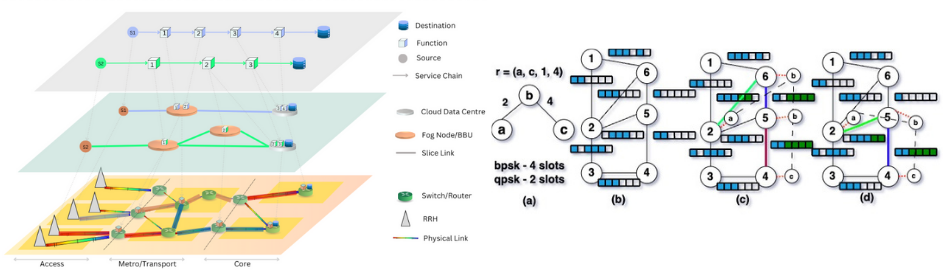}
\caption{E2E network slicing and lightpath provisioning.}
\label{netSlice}
\vspace{-1 em}
\end{figure*}

\section{Background Literature}

In single-core EONs, several studies formulated the VNE problem in conjunction with routing and spectrum allocation (RSA). Gong and Zhu~\cite{6679238} introduced one of the earliest formulations for Virtual Optical Network Embedding (VONE) over EONs, proposing both ILP and heuristic strategies based on layered-auxiliary-graph modeling. Wang and Hu~\cite{9309337} later improved scalability through a path-growing approach for optical VNE in SLICE networks, achieving near-optimal results while reducing variable complexity. However, these works primarily focused on spectrum utilization and connectivity and did not incorporate spatial division multiplexing or compute-aware placement. Similarly, cross-haul and 5G transport studies, such as Li~\textit{et al.}~\cite{8004167} and Gu~\textit{et al.}~\cite{10288372}, integrated slicing and resource allocation across RAN and metro layers, often using heuristic-assisted deep reinforcement learning, but these solutions operate at higher architectural layers and are unsuitable for fast, core-level optical provisioning.

With the emergence of multi-core fibers, research on SDM-EONs has addressed the Routing–Core–Spectrum Assignment (RCSA) and Routing–Modulation–Core–Spectrum Assignment (RMCSA) problems, frequently incorporating inter-core crosstalk (IC-XT) constraints. Zhang~\textit{et al.}~\cite{9874982} and Kumar~\textit{et al.}~\cite{9831362} proposed crosstalk-aware VONE algorithms that mitigate IC-XT in heterogeneous and spectrally-spatially elastic optical networks, respectively. Jin~\textit{et al.}~\cite{10209960} further introduced a dynamic topology-aware VONE for hybrid services in SDM-EONs, jointly considering VON topology and physical-layer topology to improve spectral efficiency. 
Tang~\textit{et al.}~\cite{Tang2020} later proposed an ILP model minimizing inter-core crosstalk jointly in spatial, frequency, and time domains for scheduled lightpath demands. More recent RMCSA heuristics, such as Heera~\textit{et al.}~\cite{heera_crootalk_rmcsa,heera_congestion_rmcsa}, introduced congestion-aware and group-processing algorithms for dynamic provisioning and fair spectrum allocation under heterogeneous traffic. While these studies effectively address crosstalk, congestion, and fairness, they treat compute resource placement as independent of optical resource allocation.

Extensive research on slice mapping and Virtual Network Embedding (VNE) in optical networks has focused on efficiently mapping virtual networks onto underlying elastic and space-division multiplexed optical infrastructures. Most optical VONE studies adopt a one-to-one node mapping for simplicity and to avoid concentrated compute contention and single-point failures. In contrast, the broader VNE community has examined node co-location as a means of improving embedding efficiency. Wang \textit{et al.}~\cite{NodeFusion} proposed the \textit{Node-Fusion} approach, which evaluates candidate substrate nodes using an integrated ranking metric based on CPU availability, node degree, and adjacent link bandwidth, and introduces a Node-Fusion Interconnection Value (NFIV) to co-locate highly interconnected virtual nodes. This reduces the number of virtual links requiring physical provisioning and minimizes overall embedding cost. Chowdhury \textit{et al.}~\cite{ChowdhuryCoordinatedVNE} further advanced this idea by coordinating node and link mapping, where node placement decisions directly consider link feasibility and bandwidth cost, enabling more balanced utilization and higher acceptance ratios. While both studies demonstrate that coordinated mapping and node co-location can significantly lower cost and improve resource efficiency, they are confined to abstract VNE models that exclude optical-layer factors such as spectrum continuity, modulation reach, and inter-core crosstalk. To the best of our knowledge, such co-location principles have not been incorporated into optical or SDM-EON slice mapping. 

Our work bridges this gap by introducing controlled node co-location within an optical slicing framework that sequentially, with balanced placement, accounts for compute constraints, spectrum allocation, and physical-layer impairments under dynamic traffic conditions. This extension preserves the efficiency benefits of Node-Fusion while adapting them to the stricter feasibility requirements of optical networks.
Adopting co-location in SDM-EONs could further lower spectrum usage (fewer inter-node virtual links) at the expense of higher per-node compute demands and potentially increased risk concentration. While two baselines (Sorted Approach and Greedy Approach) follow the common one-to-one assumption for fair comparison, the framework (DPSM for baseline and WMSM as proposed approach) can be extended to permit controlled co-location by relaxing the node-mapping cardinality constraints and augmenting compute-capacity checks.

\begin{table}[htbp]
\centering
\caption{Comparison of related works by SDM-EON, coordination, and co-location characteristics}
\setlength{\tabcolsep}{2pt} 
\renewcommand{\arraystretch}{1.05} 
\begin{tabular}{p{2.7cm}ccc p{2.3cm}}
\toprule
\textbf{Reference(s)} & 
\textbf{SDM-EON} & 
\textbf{Coord.} & 
\textbf{Co-loc.} & 
\textbf{Solution Type(s)} \\
\midrule
\cite{9874982,9831362,10209960,2593479,Tang2020,heera_crootalk_rmcsa,heera_congestion_rmcsa} 
& \cmark & \xmark & \xmark & ILP, Heuristic \\
\cite{ChowdhuryCoordinatedVNE,10288372} 
& \xmark & \xmark & \cmark & RL, Heuristic \\
\cite{NodeFusion} 
& \xmark & \cmark & \cmark & Heuristic \\
\cite{6679238,9309337} 
& \xmark & \xmark & \xmark & ILP, Heuristic \\
\midrule
\textbf{WMSM} 
& \cmark & \cmark & \cmark & Heuristic \\
\bottomrule
\end{tabular}
\end{table}

\section{System Model and Problem Formulation}

We consider an SDM-EON modeled as a graph $G = (V,E)$, where $V$ is the set of physical nodes and $E$ is the set of optical fiber links. Each link supports multiple spatial cores $\Phi_e$, each offering a frequency grid of $f$ discrete spectrum slots. A subset $V_c \subseteq V$ denotes compute-capable nodes with finite processing capacities. Cores are grouped by spatial adjacency to mitigate inter-core crosstalk.

\subsection{Definitions and Operational Semantics}

\textbf{Compute Resource Model.} Compute denotes abstract processing capacity at compute-capable nodes, expressed in normalized units representing VNF execution (e.g., transcoding, caching, monitoring). Each slice request specifies an aggregate compute demand without modeling individual VNF internals.

\textbf{OEO Semantics.} Deploying a VNF at a compute-capable waypoint triggers an Optical--Electrical--Optical (OEO) conversion, terminating the incoming lightpath and initiating a new one toward the next segment. Consequently, spectrum continuity is enforced only within each segment, not across waypoints, enabling independent modulation and spectrum assignment per segment and improving feasibility under reach and fragmentation constraints.

Each slice request $r$ is represented as $(s, d, b_{req}, c_{req})$, where $s$ and $d$ are source and destination, $b_{req}$ is bandwidth (Gbps), and $c_{req}$ is the compute requirement. A request is accepted if compute and spectrum resources are jointly provisioned along a feasible path or via designated waypoints, with each allocation decision influencing the feasibility of subsequent ones.

\subsection{Cost Model}
For each slice request $r_i$, the provisioning cost is defined as
\begin{equation}
C_{r_i} = \sum_{v \in V_c} c_{v,r_i}\,\mathbb{C}_{compute} + \sum_{e \in E} \gamma_{e,r_i}\,\mathbb{C}_{spectrum},
\end{equation}
where $\mathbb{C}_{compute}$ and $\mathbb{C}_{spectrum}$ are the per-unit costs of compute and spectrum resources, $c_{v,r_i}$ is the number of compute units allocated to $r_i$ at node $v$, and $\gamma_{e,r_i}$ is the number of spectrum slots allocated to $r_i$ on link $e$. The overall objective is to minimize the total provisioning cost across all requests:
\begin{equation}
\text{Minimize:} \quad \sum_{i=1}^{N} C_{r_i}.
\end{equation}



\subsection{Constraints}

The slice mapping must satisfy the following constraints:

\begin{itemize}





\item \textbf{(C1) Modulation reach:} The total path length must satisfy
\begin{equation}
\sum_{e \in P_{r_i}^k} D_e \leq D_{\max}\left(M_{r_i}^k\right),
\quad \forall r_i,; \forall k
\end{equation}
where $P_{r_i}$ is the selected path, $D_e$ is the distance of link $e$, and $D_{\max}(M_{r_i})$ is the maximum transmission reach for modulation format $M_{r_i}$.

\item \textbf{(C2) Spectrum contiguity:} Each request $r_i$ occupies a contiguous slot set 
\begin{equation}
\Gamma_{r_i,e,\phi} = {s_{r_i}, s_{r_i}+1, \dots, s_{r_i}+f_{r_i}-1},
\quad \forall r_i,; \forall e \in P_{r_i}^k
\end{equation}
where
\begin{equation}
f_{r_i} = b_{req} \times \eta(M_{r_i})
\end{equation}
is the number of slots per Gbps for the chosen modulation format.

\item \textbf{(C3) Spectrum continuity:} Identical contiguous slots are assigned on all links along the path, i.e., 
\begin{equation}
\Gamma_{r_i,e,\phi} = \Gamma_{r_i,e',\phi},
\quad \forall e,e' \in P_{r_i} 
\end{equation}

Spectrum continuity across different segments 
$k$ is not required due to the OEO conversion at the compute waypoints.

\item \textbf{(C4) Compute feasibility:} The allocated compute resources must satisfy 
\begin{equation}
\sum_{v \in V_c} c_{v,r_i} \geq c_{req},
\quad \forall r_i, \quad |V_c| \geq 1
\end{equation}
\begin{equation}
\sum_{r_i} c_{v,r_i} \leq C^{\text{comp}}_{v},
\quad \forall v \in V_c
\end{equation}
where $c_{v,r_i}$ is the compute assigned to request $r_i$ at node $v$ and $C^{\text{comp}}_{v}$ is the node capacity.

\item \textbf{(C5) Core-group selection (crosstalk avoidance):} Each request is assigned a core $\phi_{r_i} \in \mathcal{G}_k$, where $\mathcal{G}_k$ denotes a non-adjacent core group (e.g., $\mathcal{G}_1=\{1,3,5\}$, $\mathcal{G}_2=\{2,4,6\}$, $\mathcal{G}_3=\{7\}$), ensuring $\sum_k g_{r_i,k}=1$ with $g_{r_i,k}=1$ if $\mathcal{G}_k$ is used by $r_i$ and $0$ otherwise. This avoids assigning adjacent cores simultaneously, minimizing inter-core crosstalk.

\end{itemize}

A request is blocked if any of the above constraints are violated.

\subsection{Performance Metrics}
The following metrics are used to evaluate performance:
\begin{itemize}
    \item \textit{Blocking ratio:} fraction of slice requests rejected.
    \item \textit{Spectrum utilization:} 
    \begin{equation}
    U = \frac{\sum_{i=1}^{N}\sum_{e \in E} \gamma_{e,r_i}}{\mathbb{N}},
    \end{equation}
    where $\mathbb{N}$ is the total number of slots in the network (links $\times$ cores $\times$ slots per core).
    \item \textit{Provisioning cost:} average resource cost across accepted requests.
\end{itemize}

\subsection{Modeling Assumptions}

The slice-embedding problem, combining RMCSA with compute placement, is NP-hard, generalizing Routing and Spectrum Assignment (RSA)~\cite{ChatterjeeTutorial}, Virtual Network Embedding (VNE)~\cite{ChowdhurySurveyVNE}, and multidimensional bin-packing under modulation reach, crosstalk, and compute capacity constraints~\cite{BianzinoSurveyNP}. Exact optimization is therefore intractable, motivating the sequential heuristic framework proposed here, which integrates compute placement with spectrum allocation while preserving interdependencies among routing, compute assignment, and spectrum decisions. Two approaches are evaluated: a baseline \textit{Direct Path-Driven Slice Mapping (DPSM)} and the proposed \textit{Waypoint-Assisted Multi-Segment Slice Mapping (WMSM)}.

Requests may consolidate or distribute virtual functions across nodes following flexible VNF placement models~\cite{8543228}. Although the framework supports multiple waypoints, this study restricts to at most two for computational tractability, extending prior RMCSA formulations to compute-aware slice mapping under realistic constraints.

\section{Proposed Solution}

We propose a sequential yet integrated heuristic framework for compute-aware slice mapping in SDM-EONs, coupling routing, modulation, core selection, and spectrum allocation with compute placement. Two strategies are examined: a baseline \textit{Direct Path-Driven Slice Mapping (DPSM)} and the proposed \textit{Waypoint-Assisted Multi-Segment Slice Mapping (WMSM)}, with feasibility constraints propagating interdependently across all mapping stages.

\subsection{Direct Path-Driven Slice Mapping (DPSM)}

DPSM follows a straightforward end-to-end mapping procedure. Given a slice request $(s, d, BW, C_{req})$, the algorithm computes $k=3$ shortest paths using Dijkstra's algorithm~\cite{Tang2020}. For each candidate path $p_i$, compute feasibility by checking $\sum_{v \in p_i \cap V_c} C_v \geq C_{req}$; if satisfied, compute resources are mapped onto the first node along $p_i$ individually meeting $C_{req}$. The modulation format is then selected based on total path length $D_{p_i}$: 16-QAM for $D_{p_i} \leq 500$~km, QPSK for $D_{p_i} \leq 1000$~km, and BPSK otherwise. The required slot count $S = BW \times \eta(M)$ is computed, a core group $\{\mathcal{G}_k\}$ is chosen by the highest available slot ratio, and a contiguous block of $S$ slots is allocated on the first eligible core. If allocation fails, the next path is tried; if all $k$ paths fail, the request is blocked. Among feasible paths, the shortest is selected.

DPSM's end-to-end constraint forces a single modulation format over the full path, often requiring BPSK and consuming more slots. Its greedy compute placement also risks hotspots under high load, making it a useful but conservative benchmark for WMSM.

\subsubsection{DPSM with Balanced Placement (DPSM-B)}

DPSM-B replaces the greedy compute placement with a segment-length balancing strategy. Since OEO conversion at the compute node creates two independent segments $(s, v^*)$ and $(v^*, d)$ with independently assigned modulation formats, placing the compute node near the path midpoint maximizes the likelihood that both segments qualify for higher-order modulation. For each compute-capable node $v$ along $p_i$, the segment imbalance is:
\begin{equation}
    \delta_v = \left| d(s, v) - d(v, d) \right|,
\end{equation}

\begin{algorithm}[H]
\caption {Direct Path-Driven Slice Mapping with Balanced placement (DPSM-B)}
\begin{algorithmic}
\State \textbf{Input:} Slice Request $(s, d, BW, C_{req})$

\State Find $k = 3$ shortest paths from $s$ to $d$ using Dijkstra's algorithm

\State $feasible \gets$ \textbf{false}
\For {each path $p_i$ in order}
    \State Calculate total compute capacity along $p_i$
    \If {capacity $\geq C_{req}$}
        \For {each candidate compute node $v$ along $p_i$}
            \State Compute $\delta_v = \left| d(s, v) - d(v, d) \right|$
        \EndFor
        \State Select node(s) $v^*$ minimizing $\delta_v$
        \If {multiple nodes needed}
            \State Select combination minimizing max segment length disparity
        \EndIf
        \State Map compute resources onto $v^*$ to satisfy $C_{req}$

        \State \textit{// Spectrum checked independently per segment}
        \State Select modulation format for segment $(s, v^*)$ based on $D_1 = d(s, v^*)$
        \State Compute required slots $S_1 = BW \times$ (slots per Gbps) for $(s, v^*)$
        \State Select core group for $(s, v^*)$ with highest available slot ratio
        \For {each core $\phi_1$ in selected group}
            \If {$\phi_1$ has $S_1$ contiguous free slots}
                \State Allocate slots on $\phi_1$
                \State \textbf{break}
            \EndIf
        \EndFor

        \State Select modulation format for segment $(v^*, d)$ based on $D_2 = d(v^*, d)$
        \State Compute required slots $S_2 = BW \times$ (slots per Gbps) for $(v^*, d)$
        \State Select core group for $(v^*, d)$ with highest available slot ratio
        \For {each core $\phi_2$ in selected group}
            \If {$\phi_2$ has $S_2$ contiguous free slots}
                \State Allocate slots on $\phi_2$
                \State \textbf{break}
            \EndIf
        \EndFor

        \If {both segments allocated successfully}
            \State $feasible \gets$ \textbf{true}
            \State \textbf{break}
        \EndIf
    \EndIf
\EndFor

\If {$feasible =$ \textbf{false}}
    \State \textit{Fallback:} execute standard DPSM
    \State \textbf{return}
\EndIf

\State Update metrics: blocking rate, utilization, cost
\end{algorithmic}
\end{algorithm}

\begin{algorithm}[H]
\caption {Waypoint-Assisted Multi-Segment Slice Mapping with Balanced PLacement (WMSM-B)}
\begin{algorithmic}
\State \textbf{Input:} Slice request $(s, d, BW, C_{req})$
\State \textbf{Output:} Mapping decision (accepted/rejected), spectrum and compute allocation

\State Identify a candidate set of compute-capable waypoints $\mathcal{W}$ such that total available capacity $\sum_{w_i \in \mathcal{W}} C_{w_i} \geq C_{req}$

\State For each candidate waypoint configuration, compute segment balance:
\[
\begin{aligned}
\delta_{\mathcal{W}} &= \max_{i} \, d(w_i, w_{i+1}) - \min_{i} \, d(w_i, w_{i+1}), \\
&\quad \text{where } w_0=s, \, w_{|\mathcal{W}|+1}=d
\end{aligned}
\]
\State Select waypoint(s) $\mathcal{W}^*$ that jointly minimize total end-to-end distance and segment imbalance:
\[
\begin{aligned}
\min_{\mathcal{W}} \left( \sum_{i=0}^{|\mathcal{W}|} d(w_i, w_{i+1}) + \lambda \cdot \delta_{\mathcal{W}} \right), \\
\quad \text{where } w_0=s, \, w_{|\mathcal{W}|+1}=d
\end{aligned}
\]
\Comment{For simplicity, we consider up to two waypoints in this paper}

\For{each segment between consecutive nodes $(w_i, w_{i+1})$}
    \State Compute up to $k=2$ shortest paths
    \State Estimate distance $D_{i}$ for each path
    \State Select feasible modulation format based on $D_{i}$
    \State Compute required slots $S_{i} = BW \times (\text{slots per Gbps})$
\EndFor

\State Combine feasible segment paths to form candidate routes
\State Choose the route minimizing total spectrum usage $\sum_i S_{i}$

\For{each segment in the selected route}
    \For{each core $\phi$ in the assigned core group}
        \If {$\phi$ has $S_i$ contiguous free slots}
            \State Allocate $S_i$ slots on $\phi$
            \State \textbf{break}
        \EndIf
    \EndFor
    \If {no allocation is feasible}
        \State \textit{Fallback:} execute standard WMSM
        \State \textbf{return}
    \EndIf
\EndFor

\For{each selected waypoint $w_i \in \mathcal{W}^*$}
    \State Allocate compute capacity $C_{req,i}$ ensuring $\sum_i C_{req,i} = C_{req}$
\EndFor

\State Update network metrics: blocking rate, utilization, cost
\end{algorithmic}
\end{algorithm}

and $v^* = \arg\min_v \delta_v$ is selected. When a single node cannot satisfy $C_{req}$, the algorithm selects the pair $(v_1, v_2)$ minimizing the maximum length among segments $(s, v_1)$, $(v_1, v_2)$, and $(v_2, d)$.

Spectrum allocation then proceeds independently per segment: modulation $M_k$ and slot count $S_k = BW \times \eta(M_k)$ are determined from segment length $D_k$, a core group is selected by the highest available slot ratio, and contiguous slots are allocated on the first eligible core. If either segment fails, the algorithm advances to path $p_{i+1}$; if all $k$ paths are exhausted, it falls back to standard DPSM.


\subsection{Waypoint-Assisted Multi-Segment Slice Mapping (WMSM)}

Unlike DPSM, WMSM introduces intermediate compute-capable waypoints that decompose each slice request into shorter optical segments. OEO conversion at each waypoint relaxes spectrum continuity to within-segment only, enabling independent modulation format selection per segment and reducing total slot consumption.

Given a request $(s, d, BW, C_{req})$, WMSM enumerates candidate waypoint configurations $\mathcal{W} \subseteq V_c$ satisfying $\sum_{w_i \in \mathcal{W}} C_{w_i} \geq C_{req}$, restricted to at most two waypoints for tractability. For each configuration, up to $k=2$ shortest paths are computed per consecutive segment $(w_i, w_{i+1})$. Modulation is assigned per segment length $D_i$ (16-QAM: $\leq$500~km, QPSK: $\leq$1000~km, BPSK: otherwise), and slot count $S_i = BW \times \eta(M_i)$ is derived accordingly. The configuration minimizing $\sum_i S_i$ is selected as $\mathcal{W}^*$.

Spectrum allocation then proceeds segment by segment: for each $(w_i, w_{i+1})$, a core group is chosen by the highest available slot ratio, and the first contiguous block of $S_i$ free slots is allocated. If any segment fails, the configuration is discarded, and the next candidate is evaluated; if all configurations fail, the request is blocked.

 Upon successful allocation across all segments, compute capacity is distributed across the selected waypoint nodes. The waypoint choice determines segment lengths, which dictate modulation and slot requirements, which in turn govern spectrum feasibility. If allocation fails, the algorithm backtracks to the next-best configuration, achieving compute--spectrum coordination without exhaustive joint optimization.

\subsubsection{WMSM with Balanced Placement (WMSM-B)}

WMSM-B augments waypoint selection with a segment-length balance objective, since unbalanced segments can force lower-order modulation on longer segments, undermining spectral efficiency. For each candidate configuration $\mathcal{W}$, the imbalance is:
\begin{equation}
    \delta_{\mathcal{W}} = \max_{i}\, d(w_i, w_{i+1}) - \min_{i}\, d(w_i, w_{i+1}),
\end{equation}
and the selection objective becomes:
\begin{equation}
    \mathcal{W}^* = \arg\min_{\mathcal{W}} \left( \sum_{i=0}^{|\mathcal{W}|} 
    d(w_i, w_{i+1}) + \lambda \cdot \delta_{\mathcal{W}} \right),
\end{equation}
where $\lambda > 0$ trades off total path length against segment balance. Per-segment modulation, slot allocation, and fallback proceed identically to baseline WMSM. If allocation fails, the algorithm backtracks through configurations ranked by the augmented objective until a feasible mapping is found or all options are exhausted.

Both WMSM and WMSM-B enforce spectrum continuity, contiguity, and non-overlap within each segment, with inter-core crosstalk mitigated via group-based core selection. For example, a 1500~km path requiring QPSK end-to-end can be split into 700~km and 800~km segments, both feasible under 16-QAM, halving the spectrum cost. WMSM's complexity scales polynomially with network size and linearly with candidate waypoints, making it practical for dynamic SDM-EONs.

\subsection{Baseline Approaches}

\subsubsection{Sorted Approach}

Inspired by the strategy proposed in \cite{10209960}, we implement a baseline approach: the request prioritization mechanism, where slice requests are first sorted based on their compute and bandwidth demands. Prior to resource allocation, physical nodes are ranked by available compute capacity, while adjacent links are ordered according to spectrum slot availability. Unlike \cite{10209960}, which considers virtual topology cycles and maps them onto matching physical cycles, our implementation simplifies the scenario by treating each request as an aggregate compute demand without requiring cycle-aware embedding.

\subsubsection{Greedy Approach}

A greedy baseline approach is also implemented, where slice requests are provisioned using a first-come, first-served (FCFS) strategy. For compute resource allocation, the first physical node encountered along the candidate path that satisfies the compute requirement is selected. Similarly, for spectrum allocation, the first available core on each traversed link that can accommodate the required contiguous spectrum slots is chosen, without global optimization across requests.

\section{Numerical Evaluation}

\subsection{Simulation Setup}

We consider bandwidth demands of 1–20 Gbps and compute needs of 5–10 units, representing light to moderate workloads such as transcoding, caching, and monitoring \cite{8647700}. These bounds correspond to 1–20 spectrum slots depending on modulation, consistent with prior RMCSA traffic models \cite{Tang2020}. 

We evaluate our proposed slice mapping approaches on the 14-node NSF network topology, a widely used reference in optical networking research \cite{9831362}. Each bidirectional link consists of 7 cores, with 120 frequency slots per core. Link distances range from 350~km to 1500~km. Three modulation formats are supported: BPSK (2000~km reach, 4~slots/Gbps), QPSK (1000~km, 2~slots/Gbps), and 16-QAM (500~km, 1~slot/Gbps). Spectrum slots are assigned a unit cost of 40 per Gbps, while compute resources cost 400 units per unit capacity.

To focus on spectrum effects, compute capacities are initially set high (4000 units per node). In subsequent experiments, compute capacities are reduced (400 units per node) to evaluate the impact of coupled compute–spectrum constraints. Eleven nodes are randomly selected as compute-capable.

Slice requests arrive according to a Poisson process with exponentially distributed holding times (mean 0.5~hours). Each experiment generates 50{,}000 requests between random source–destination pairs. Bandwidth demands are uniformly distributed between 1–20~Gbps, and compute requirements are drawn uniformly from 5–10 units. The selected range of 1–20~Gbps ensures that even the largest requests remain feasible under BPSK modulation, which requires up to 4~slots/Gbps. For instance, a 20~Gbps request under BPSK would occupy 80~slots, remaining within the 120~available slots per core per link. This range thus allows fair comparison across modulation formats while avoiding infeasible allocations. We also conducted experiments with bandwidths up to 40~Gbps and observed consistent performance trends; however, these results are omitted here due to page-length constraints. For routing, we use $k=3$ shortest paths in DPSM and $k=2$ per segment in WMSM. Each data point is averaged over 30 independent runs, and the observed variance is below 5\%, ensuring statistical reliability.

We have evaluated the proposed approaches and the baseline approaches with respect to blocking rate, spectrum utilization, and provisioning cost as performance metrics. Prior works on RMCSA \cite{heera_crootalk_rmcsa} have demonstrated that algorithms similar to WMSM reduce fragmentation and improve fairness, which aligns with our observations.

We compare our approaches against a baseline prioritization strategy inspired by \cite{10209960}, where requests are sorted by compute and bandwidth requirements before allocation. Nodes are ranked by available compute capacity and adjacent links by spectrum availability. Unlike cycle-aware embedding in \cite{10209960}, our baseline simplifies requests as aggregate compute demands without virtual topology cycles. This provides a conservative benchmark to evaluate the relative improvements of DPSM and WMSM.

\begin{figure}[!t]
\centering

\begin{minipage}{\linewidth}
  \includegraphics[width=\linewidth]{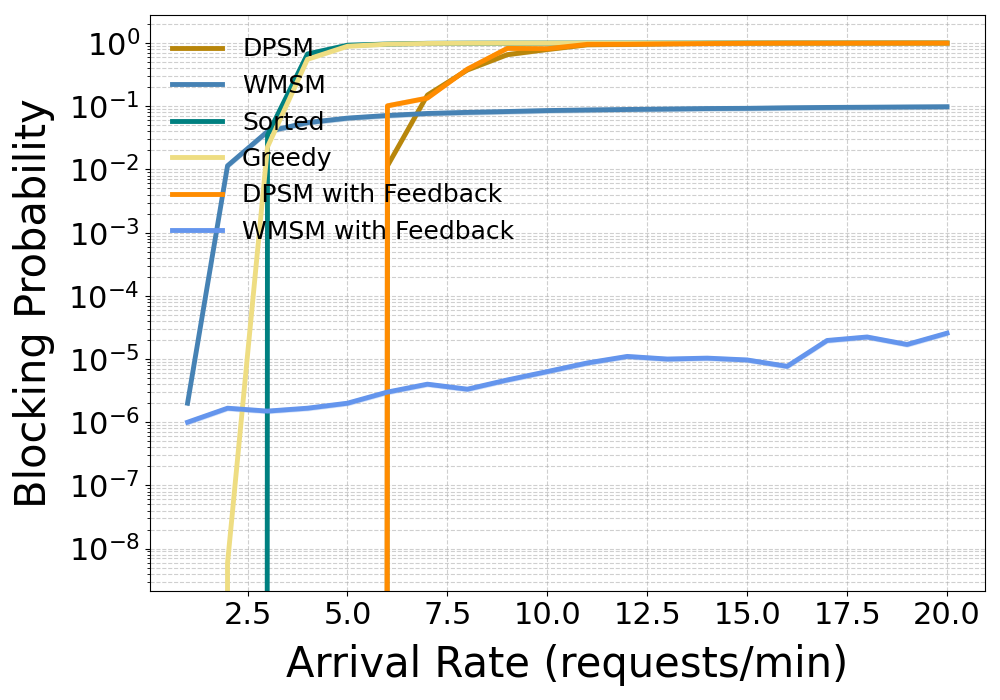}
  \caption{Blocking Rate vs. Arrival Rate.}
\end{minipage}
\hfill

\vspace{0.3cm}

\begin{minipage}{\linewidth}
  \includegraphics[width=\linewidth]{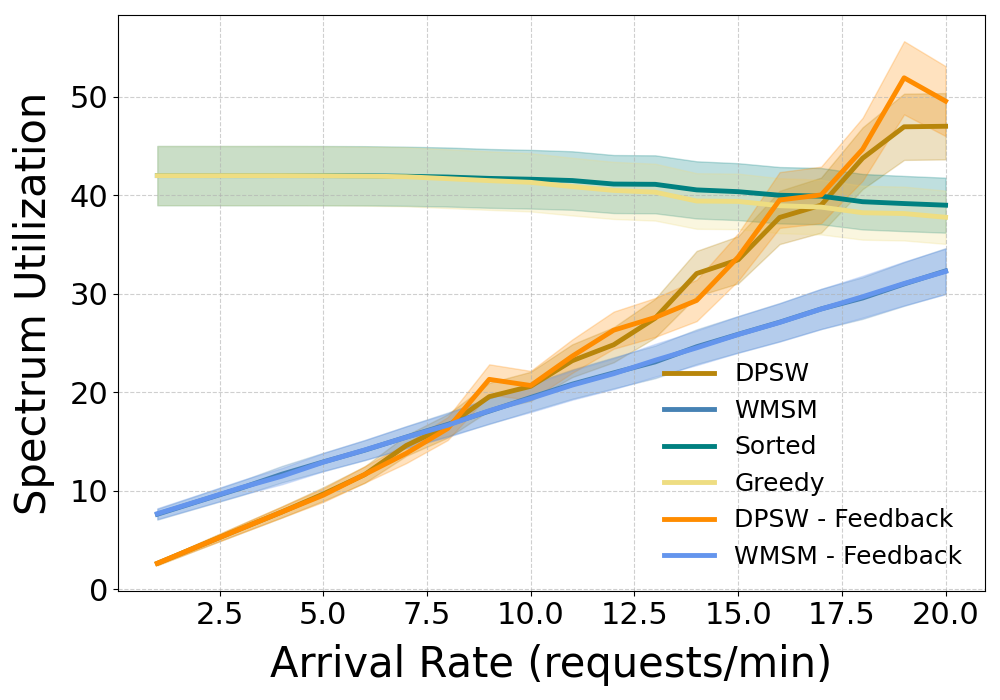}
  \caption{Spectrum Utilization vs. Arrival Rate.}
\end{minipage}
\hfill


\begin{minipage}{\linewidth}
  \includegraphics[width=\linewidth]{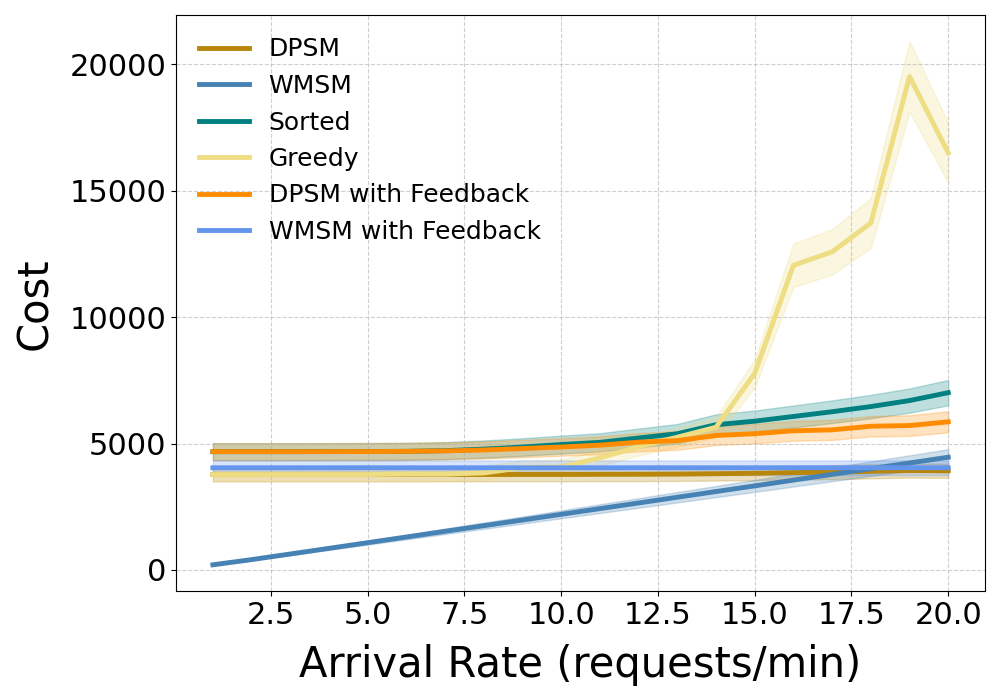}
  \caption{Provisioning Cost vs. Arrival Rate.}
\end{minipage}
\hfill

\vspace{-0.2cm}
\end{figure}

\begin{figure}[!t]
\centering

\begin{minipage}{\linewidth}
  \includegraphics[width=\linewidth]{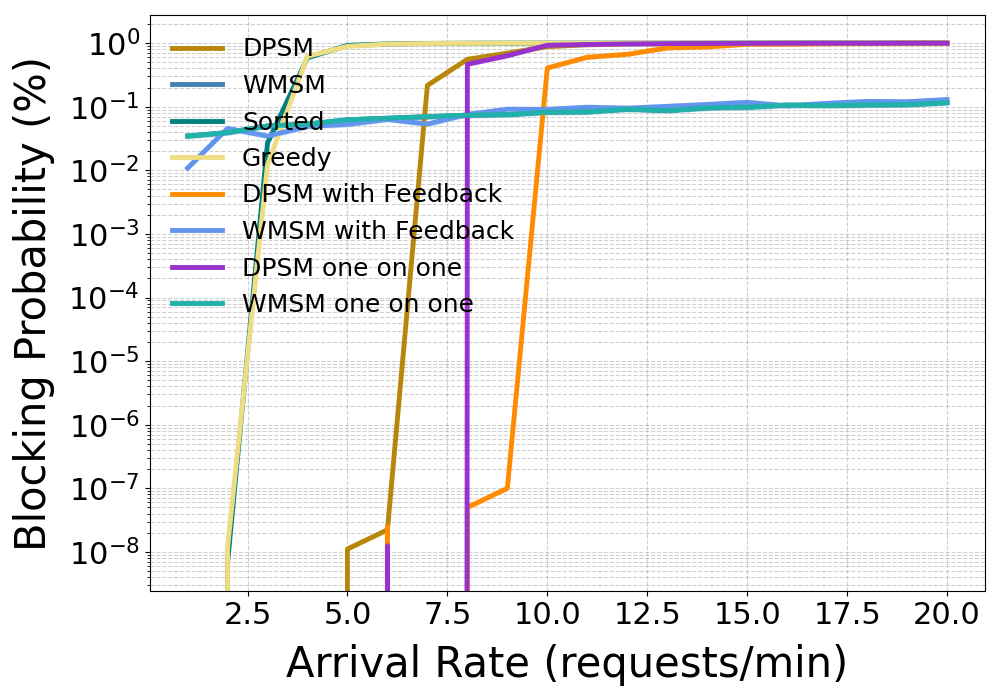}
  \caption{Blocking Rate vs. Arrival Rate for limited compute capacity.}
\end{minipage}

\vspace{0.3cm}

\begin{minipage}{\linewidth}
  \includegraphics[width=\linewidth]{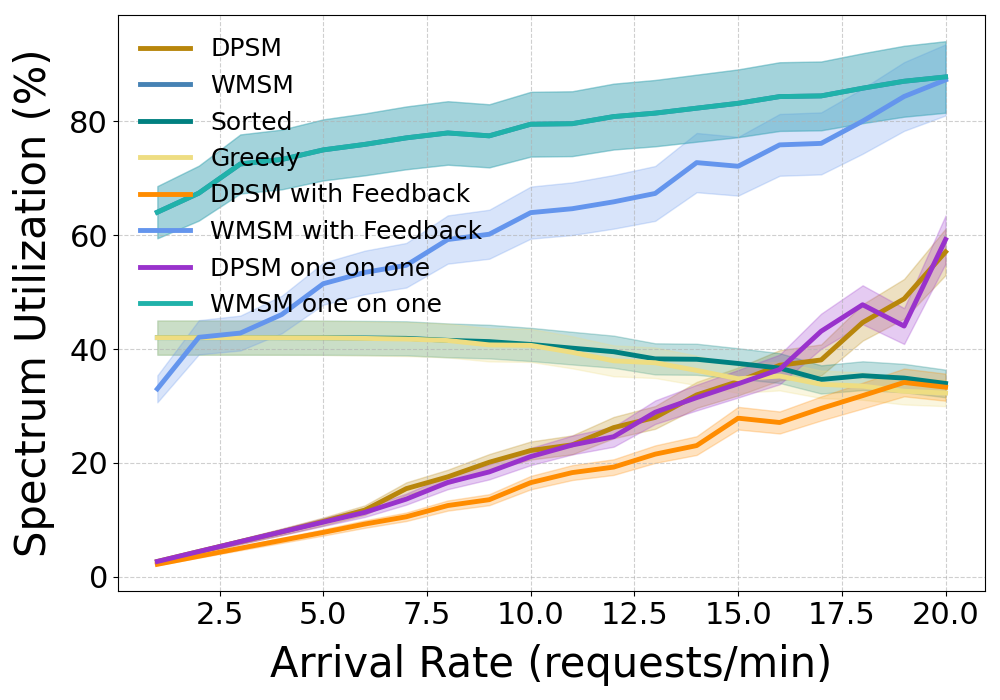}
  \caption{Spectrum Utilization vs. Arrival Rate for limited compute capacity.}
\end{minipage}


\begin{minipage}{\linewidth}
  \includegraphics[width=\linewidth]{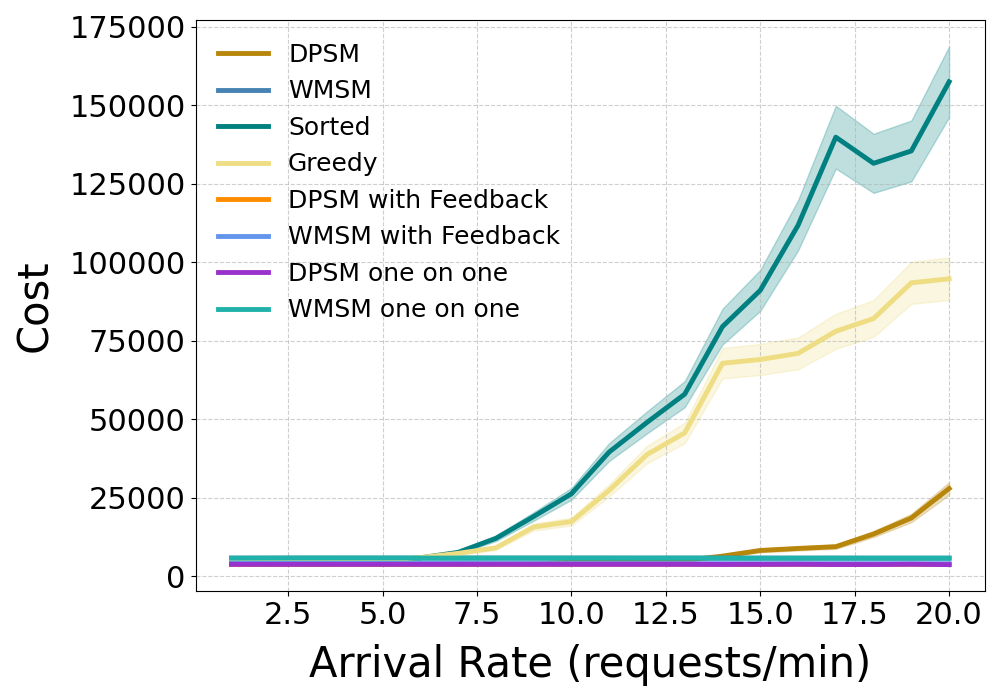}
  \caption{Provisioning Cost vs. Arrival Rate for limited compute capacity.}
\end{minipage}

\vspace{-0.2cm}
\end{figure}


\subsection{Results and Discussion}
Figures 2, 3, and 4 show blocking rate, spectrum utilization, and cost under high compute availability (4000 units per node). Figures 5, 6, and 7 present the same metrics when compute capacity is limited to 400 units per node. 

Approaches compared:
\begin{itemize}
    \item \textbf{One-on-one:} Maps each virtual node to a distinct physical node without fusion, serving as a baseline to isolate the benefit of node fusion, where merging multiple virtual nodes onto a single physical node reduces resource consumption and improves scalability.

    \item \textbf{With Balanced Placement:} Integrates VONE and RMCSA sequentially with a balanced placement mechanism, where spectrum allocation outcomes are informed with refined node mapping decisions, unlike conventional decoupled approaches that solve the two stages independently.

    \item \textbf{Sorted and Greedy:} Standard baseline heuristics for RMCSA that assign spectrum slots in sorted or greedy order without any awareness of virtual topology or waypoint structure, included to benchmark the gain from our proposed decomposition strategies.

\end{itemize}

The results highlight three key insights:
\begin{itemize}
    \item \textbf{Blocking performance:} WMSM with Feedback achieves strictly the lowest 
    blocking probability across all arrival rates, particularly at higher loads, owing to its 
    balanced segment placement and independent per-segment spectrum allocation. Notably, 
    WMSM with Feedback's blocking rate continues to decline beyond the plotted range, 
    plateauing only at much higher arrival rates (around 50--60 requests/min), demonstrating 
    that the scheme possesses significant untapped capacity headroom well beyond the 
    conditions evaluated here. Standard WMSM also sustains comparably low blocking, and both 
    WMSM and WMSM one-on-one exhibit nearly identical blocking behavior, a direct 
    consequence of their similar waypoint selection strategies, which yield equivalent segment 
    decompositions in practice. It is worth noting that WMSM one-on-one was implemented 
    exclusively under limited compute capacity, as the effect of one-on-one processing is 
    most meaningful and observable when resources are constrained. DPSM with Feedback 
    sustains near-zero blocking until moderate arrival rates, beyond which it rises sharply, 
    while standard DPSM stabilizes at higher blocking probabilities from early arrival rates 
    onward. Both Sorted and Greedy benchmarks saturate quickly, confirming their inability to 
    handle dynamic traffic conditions compared to the proposed approaches.

    Under limited capacity, blocking behavior shifts significantly. WMSM with Feedback 
    sustains strictly the lowest blocking across all arrival rates, with WMSM and WMSM 
    one-on-one again performing near-identically and close behind. DPSM one-on-one and 
    standard DPSM saturate at moderate arrival rates, while Sorted and Greedy collapse even 
    earlier, confirming that capacity constraints disproportionately impact less adaptive 
    schemes.

    \item \textbf{Spectrum utilization:} Sorted and Greedy maintain high but declining 
    utilization across all arrival rates as they saturate early. DPSM and DPSM with Feedback 
    grow steadily from low utilization, with DPSM with Feedback exhibiting higher variance 
    at peak loads. WMSM with Feedback consistently achieves the lowest utilization, growing 
    gradually and confirming that feedback-driven waypoint placement yields superior spectrum 
    efficiency across all traffic intensities. WMSM and WMSM one-on-one track each other 
    closely in utilization, reinforcing that the two waypoint selection strategies converge 
    to similar resource consumption patterns, both remaining significantly more efficient 
    than DPSM-based schemes.

    Under limited capacity, Sorted and Greedy plateau before declining as blocking increases. 
    DPSM and DPSM one-on-one grow aggressively with high variance at peak loads, indicating 
    inefficient spectrum usage under congestion. WMSM with Feedback remains the most 
    efficient throughout, with WMSM and WMSM one-on-one closely following, demonstrating 
    that multi-segment decomposition retains its spectrum efficiency advantage even under 
    constrained resources.

    \item \textbf{Provisioning cost:} WMSM with Feedback achieves the flattest cost curve, 
    minimizing spectrum and compute usage even at high traffic loads, while Sorted and Greedy 
    incur rapidly escalating costs as arrival rates grow. DPSM rises moderately, whereas 
    WMSM-based schemes maintain near-flat cost profiles across all arrival rates. The close 
    cost trajectories of WMSM and WMSM one-on-one further corroborate that their underlying 
    waypoint mechanisms produce equivalent provisioning behavior. Similarly, DPSM one-on-one 
    was evaluated only under limited compute capacity, where its impact on cost relative to 
    standard DPSM is most apparent.

    Under limited capacity, cost disparities become more pronounced. Sorted and Greedy incur 
    the steepest cost increases, while DPSM rises moderately. WMSM with Feedback maintains 
    the flattest cost profile across all arrival rates, with WMSM and WMSM one-on-one closely 
    matching it, confirming that multi-segment decomposition provides the most cost-stable 
    provisioning under capacity-constrained conditions.

\end{itemize}

Although fragmentation and fairness results are not plotted, our experiments indicate that 
WMSM exhibits lower fragmentation than DPSM, as it balances allocations across cores. This 
behavior is consistent with group-processing RMCSA studies, which emphasize fairness 
improvements from prioritization mechanisms.

The numerical evaluation presented in this section is conducted using a single reference topology, namely the NSFNET network. While results are not explicitly reported for multiple topologies, the proposed WMSM framework is not tailored to any topology-specific structural property of NSFNET. Instead, its decision-making relies on localized mechanisms, including shortest-path computation, candidate waypoint selection based on available compute capacity, and segment-level spectrum feasibility checks. The effectiveness of WMSM is therefore expected to extend to other widely used backbone topologies such as USNET or COST239. As a result, the benefits of segmentation, namely improved modulation feasibility and reduced spectrum occupancy, are preserved across networks with different diameters, node degrees, and link lengths.

\medskip
In summary, DPSM provides lightweight efficiency in balanced scenarios, while WMSM achieves superior performance under high traffic loads by trading additional transceivers for spectrum savings. Both approaches outperform the baseline, demonstrating the benefit of compute-aware slice mapping in SDM-EONs.

\section{Conclusion}

This paper presented two spectrum allocation approaches for compute-aware SDM-EON provisioning: Direct Path-Driven Slice Mapping (DPSM) and Waypoint-Assisted Multi-Segment Slice Mapping (WMSM). WMSM consistently outperforms DPSM and both Sorted and Greedy baselines across blocking probability, spectrum utilization, and provisioning cost, owing to its segment-by-segment optimization and dynamic compute node placement. The integration of a balanced placement mechanism — coupling VONE and RMCSA sequentially — further improves performance, with WMSM with Balanced placement sustaining blocking probabilities as low as $10^{-5}$ even under limited capacity. Comparisons against one-on-one node mapping confirm that node fusion yields meaningful reductions in resource consumption and cost. Under unlimited capacity, WMSM and its balanced placement variant maintain near-flat cost profiles while Sorted and Greedy incur rapidly escalating costs at high arrival rates. Under limited capacity, these disparities become more pronounced, validating the robustness of the proposed approaches under constrained resources. These findings provide actionable insights for designing flexible, compute-aware optical networks. Future work includes extending the models to composite compute function mapping, integrating compute and spectrum availability metrics into provisioning decisions, addressing fragmentation as an explicit optimization goal, and exploring optimal solutions through Reinforcement Learning approaches.

\section*{Acknowledgment}

This work was supported in part by the National Science Foundation under Grant No. CNS-2008856.

\bibliographystyle{IEEEtran}
\bibliography{reference.bib}

@ARTICLE{9874982,
  author={Zhang, Qihan and Zhang, Xu and Gong, Xiaoxue and Guo, Lei},
  journal={Journal of Lightwave Technology}, 
  title={Crosstalk-Avoid Virtual Optical Network Embedding Over Elastic Optical Networks With Heterogeneous Multi-Core Fibers}, 
  year={2022},
  volume={40},
  number={24},
  pages={7687-7700},
  doi={10.1109/JLT.2022.3203861}}

@ARTICLE{9309337,
  author={Wang, Yang and Hu, Qian},
  journal={Journal of Lightwave Technology}, 
  title={A Path Growing Approach to Optical Virtual Network Embedding in SLICE Networks}, 
  year={2021},
  volume={39},
  number={8},
  pages={2253-2262},
  doi={10.1109/JLT.2020.3047713}}

@INPROCEEDINGS{10209960,
  author={Jin, Tianyu and Yin, Shan and Huang, Shanguo},
  booktitle={2023 Opto-Electronics and Communications Conference (OECC)}, 
  title={A Dynamic VONE Algorithm Considering Topology For Hybrid Services in SDM-EON}, 
  year={2023},
  volume={},
  number={},
  pages={1-6},
  doi={10.1109/OECC56963.2023.10209960}}

@INPROCEEDINGS{9831362,
  author={Kumar, Vinay and Halder, Joy and Mitra, Abhijit and Oki, Eiji and Chatterjee, Bijoy Chand},
  booktitle={2022 IEEE 23rd International Conference on High Performance Switching and Routing (HPSR)}, 
  title={Inter-Core and Inter-Mode Crosstalk-Avoided Virtual Network Embedding in Spectrally-Spatially Elastic Optical Networks}, 
  year={2022},
  volume={},
  number={},
  pages={125-130},
  doi={10.1109/HPSR54439.2022.9831362}}

@ARTICLE{10288372,
  author={Gu, Jiahua and Zhu, Min and Wang, Yunwu and Cai, Xiaofeng and Cai, Yuancheng and Zhang, Jiao and Lei, Mingzheng and Hua, Bingchang and Gu, Pingping and Zhao, Guo},
  journal={Journal of Optical Communications and Networking}, 
  title={Resource-efficient and QoS guaranteed 5G RAN slice migration in elastic metro aggregation networks using heuristic-assisted deep reinforcement learning}, 
  year={2023},
  volume={15},
  number={11},
  pages={854-870},
  doi={10.1364/JOCN.496733}}

@ARTICLE{6679238,
  author={L. Gong and Z. Zhu},
  journal={Journal of Lightwave Technology}, 
  title={Virtual Optical Network Embedding (VONE) Over Elastic Optical Networks}, 
  year={2014},
  month={Feb.},
  volume={32},
  number={3},
  pages={450-460},
  doi={10.1109/JLT.2013.2294389}}

@article{Richardson,
author = {Richardson, D.J. and Fini, John and Nelson, Lynn},
year = {2013},
month = {05},
pages = {354-362},
title = {Space Division Multiplexing in Optical Fibres},
volume = {7},
journal = {Nature Photonics},
doi = {10.1038/nphoton.2013.94}
}

@article{2593479,
author = {Zhang, L. and Ansari, N. and Khreishah, A.},
year = {2016},
pages = {1983-1986},
title = {Anycast Planning in Space Division Multiplexing Elastic Optical Networks with Multi-Core Fibers},
volume = {20},
journal = {IEEE Communications Letters},
doi = {10.1109/LCOMM.2016.2593479}
}

@ARTICLE{8004167,
  author={Li, Xi and Casellas, Ramon and Landi, Giada and de la Oliva, Antonio and Costa-Perez, Xavier and Garcia-Saavedra, Andres and Deiss, Thomas and Cominardi, Luca and Vilalta, Ricard},
  journal={IEEE Communications Magazine}, 
  title={5G-Crosshaul Network Slicing: Enabling Multi-Tenancy in Mobile Transport Networks}, 
  year={2017},
  volume={55},
  number={8},
  pages={128-137},
  doi={10.1109/MCOM.2017.1600921}}

@ARTICLE{8543228,
  author={Yin, Shan and Huang, Shanguo and Liu, Hao and Guo, Bingli and Gao, Tao and Li, Wenzhe},
  journal={IEEE Access}, 
  title={Survivable Multipath Virtual Network Embedding Against Multiple Failures for SDN/NFV}, 
  year={2018},
  volume={6},
  number={},
  pages={76909-76923},
  doi={10.1109/ACCESS.2018.2882793}}

@INPROCEEDINGS{8647700,
  author={Li, Junling and Shi, Weisen and Ye, Qiang and Zhuang, Weihua and Shen, Xuemin and Li, Xu},
  booktitle={2018 IEEE Global Communications Conference (GLOBECOM)}, 
  title={Online Joint VNF Chain Composition and Embedding for 5G Networks}, 
  year={2018},
  volume={},
  number={},
  pages={1-6},
  doi={10.1109/GLOCOM.2018.8647700}}

@ARTICLE{heera_crootalk_rmcsa,
  author={Heera, Baljinder Singh and Sharma, Anjali and Lohani1, Varsha and Singh, Yatindra Nath},
  journal={Photonic Network Communications}, 
  title={Routing, modulation, core and spectrum assignment techniques for crosstalk management in multi-core fiber based spatially multiplexed elastic optical networks.}, 
  year={2024},
  volume={48},
  pages={35-50},
  doi={https://doi.org/10.1007/s11107-024-01021-8}}

@misc{heera_congestion_rmcsa,
      title={RMCSA Algorithm for Congestion-Aware and Service Latency Aware Dynamic Service Provisioning in Software-Defined SDM-EONs}, 
      author={Baljinder Singh Heera and Shrinivas Petale and Yatindra Nath Singh and Suresh Subramaniam},
      year={2024},
      eprint={2412.10685},
      archivePrefix={arXiv},
      primaryClass={cs.NI},
      url={https://arxiv.org/abs/2412.10685}, 
}

@article{Tang2020,
  author    = {F. Tang and Y. Li and G. Shen and G. N. Rouskas},
  title     = {Minimizing inter-core crosstalk jointly in spatial, frequency, and time domains for scheduled lightpath demands in multi-core fiber-based elastic optical networks},
  journal   = {Journal of Lightwave Technology},
  volume    = {38},
  number    = {20},
  pages     = {5595--5607},
  year      = {2020},
  doi       = {10.1109/JLT.2020.3006925}
}

@article{NodeFusion,
  author    = {Desheng Wang and Weizhe Zhang and Hui He and Chuanyi Liu},
  title     = {Node-Fusion: Topology-Aware Virtual Network Embedding Algorithm for Repeatable Virtual Network Mapping over Substrate Nodes},
  journal   = {IEEE Access},
  volume    = {9},
  pages     = {142581--142592},
  year      = {2021},
  doi       = {10.1109/ACCESS.2021.3119849}
}

@inproceedings{ChowdhuryCoordinatedVNE,
  author    = {N. M. Mosharaf Kabir Chowdhury and Muntasir Raihan Rahman and Raouf Boutaba},
  title     = {Virtual Network Embedding with Coordinated Node and Link Mapping},
  booktitle = {Proceedings of IEEE INFOCOM},
  pages     = {783--791},
  year      = {2009},
  doi       = {10.1109/INFCOM.2009.5061974}
}

@article{ChatterjeeTutorial,
  author={B. C. Chatterjee and N. Sarma and E. Oki},
  title={Routing and Spectrum Allocation in Elastic Optical Networks: A Tutorial},
  journal={IEEE Communications Surveys \& Tutorials},
  volume={17},
  number={3},
  pages={1776–1800},
  year={2015},
  doi={10.1109/COMST.2015.2431731}
}

@article{ChowdhurySurveyVNE,
  author={N. M. M. K. Chowdhury and R. Boutaba},
  title={A Survey of Network Virtualization},
  journal={Computer Networks},
  volume={54},
  number={5},
  pages={862–876},
  year={2010},
  doi={10.1016/j.comnet.2009.10.017}
}

@article{BianzinoSurveyNP,
  author={A. P. Bianzino and C. Chaudet and D. Rossi and J. Rougier},
  title={A Survey of Green Networking Research},
  journal={IEEE Communications Surveys \& Tutorials},
  volume={14},
  number={1},
  pages={3–20},
  year={2012},
  note={Discusses NP-hardness of multi-resource allocation problems}
}
\end{document}